\documentclass[journal]{IEEEtran}
\usepackage{cite}

\ifCLASSINFOpdf
  \usepackage[pdftex]{graphicx}
\else
  \usepackage[dvips]{graphicx}
\fi
\usepackage[cmex10]{amsmath}
\usepackage{algorithm}
\usepackage{algpseudocode}

\usepackage{array}

\ifCLASSOPTIONcompsoc
  \usepackage[caption=false,font=normalsize,labelfont=sf,textfont=sf]{subfig}
\else
  \usepackage[caption=false,font=footnotesize]{subfig}
\fi
\usepackage{dblfloatfix}

\usepackage{booktabs}

\usepackage{url}

\usepackage{hhline}

\usepackage{float}
\usepackage{multirow}
\usepackage{enumerate}
\usepackage{amsthm}
\usepackage{nidanfloat}

\newtheoremstyle{mystyle}
  {}
  {}
  {\itshape}
  {}
  {\bfseries}
  {.}
  { }
  {}
\theoremstyle{mystyle}

\newtheorem{remark}{Remark}

\begin{document}
%
\title{Zero-bias Deep Learning Enabled Quick and Reliable Abnormality Detection in IoT}



\author{Yongxin~Liu,
        Jian~Wang,
        Jianqiang~Li,
        Shuteng Niu, and
        Houbing~Song,~\IEEEmembership{Senior Member,~IEEE} 
\thanks{Yongxin Liu, Jian Wang, Shuteng Niu and Houbing Song are with the Department of Electrical Engineering and Computer Science, Embry-Riddle Aeronautical University, Daytona Beach, FL 32114 USA}
\thanks{Jianqiang Li is with the College of Computer Science and Software Engineering, Shenzhen University, China}
\thanks{Manuscript received March 19, 2021; revised XXX. }}

\markboth{IEEE Internet of Things Journal,~Vol.~11, No.~4, December~2020}%
{Shell \MakeLowercase{\textit{et al.}}: Bare Demo of IEEEtran.cls for Journals}
%



\IEEEtitleabstractindextext{%
\begin{abstract}
Abnormality detection is essential to the performance of safety-critical and latency-constrained systems. However, as systems are becoming increasingly complicated with a large quantity of heterogeneous data, conventional statistical change point detection methods are becoming less effective and efficient. Although Deep Learning (DL) and Deep Neural Networks (DNNs) are increasingly employed to handle heterogeneous data, they still lack theoretic assurable performance and explainability. 

This paper integrates zero-bias DNN and Quickest Event Detection algorithms to provide a holistic framework for quick and reliable detection of both abnormalities and time-dependent abnormal events in Internet of Things (IoT). We first use the zero-bias dense layer to increase the explainability of DNN. We provide a solution to convert zero-bias DNN classifiers into performance assured binary abnormality detectors. Using the converted abnormality detector, we then present a sequential quickest detection scheme which provides the theoretically assured lowest abnormal event detection delay under false alarm constraints. Finally, we demonstrate the effectiveness of the framework using both massive signal records from real-world aviation communication systems and simulated data. Code and data of our work is available at \url{https://github.com/pcwhy/AbnormalityDetectionInZbDNN}

\end{abstract}

}

\maketitle

\IEEEdisplaynontitleabstractindextext

%
\IEEEpeerreviewmaketitle

\section{Introduction}
%
%
%
%
Deep Learning (DL) has reformed the ecosystem of the Internet of Things (IoT). On the one hand, they have been successfully applied in smart devices for accurate recognition of complicated inputs \cite{xu2020rf, jiang2020applying, wang2020dynamic}. On the other hand, deep learning models do not require high-quality features and reduce the time-consuming feature engineering in conventional machine learning schemes \cite{peng2018modulation,gao2019eeg}. As a representative technology within the scope of DL, Deep Neural Network (DNN) classifiers aim to use hierarchically stacked convolution layers to extract latent features to make accurate decisions.

Although DL and DNNs are successful in general purpose applications, applying DNNs in safety-critical systems requiring assured performance is still controversial. Firstly, DNNs perform well on known subjects but cannot distinguish unseen abnormal data. Abnormal signals, such as cyberattacks, are required to identify in real-time with constrained false alarms \cite{jiang2019uncertainty}. Secondly, deep neural networks lack explainability, while applications in safety-critical systems require making accurate decisions with known and explainable behaviors. Thirdly, in safety-critical systems, classifiers are supposed to evolve efficiently within a manageable behavior. The three obstacles impede the deployment of DL and DNNs in IoT of safety-critical systems. Compared with DNNs, the nearest neighbor matching algorithms naturally overcome these obstacles and gain popularity in safety-critical systems.

To address the first challenge, existing works use deep Autoencoders or Generative Adversarial Networks (GANs) to capture the latent features of the domain-specific inputs by compressing and accurately reconstructing them. However, training deep autoencoders or GAN models is even more computationally expensive than training DNN classifiers on a specific domain. Moreover, autoencoders or GAN models do not guarantee to respond in time with constrained false alarms \cite{perera2017efficient}. 

For the second problem, the eXplainable AI (XAI) has been proposed \cite{das2020opportunities}. However, most of the related works treat DNN models as Black Boxes and focus on visualizing the importance of input features, i.e., whether a DNN model is picking the right features to make decisions and do not provide insights on models' performance boundaries. In safety-critical scenarios, decision boundaries or interfaces to diagnose the error risks are more important factors for assurability. 

Finally, to support dynamic evolving DNN models, continual learning models such as Elastic Weight Consolidation (EWC \cite{kirkpatrick2017overcoming}) and Knowledge Replay (KR \cite{rolnick2019experience}) are proposed. However, knowledge replay requires training old data generators, which is computationally expensive while EWC algorithms are efficient, but they are subject to numerical stability issues.

In this paper, we utilize an enhanced deep learning framework based on our previous work \cite{liu2020zero},  the zero-bias dense layer enabled DNN, for quick and reliable detection of abnormalities with assured performance. We use zero-bias dense layers to facilitate DNNs with both non-impaired and explainable performance in distinguishing known or abnormal inputs and rapidly react to abnormalities with minimum latency and false alarm constraints. Furthermore, our solution efficiently derives abnormality detectors from existing DNN classifiers. The effectiveness of the proposed framework in handling massive signal recognition has been demonstrated. The contributions of this paper are as follows:
\begin{itemize}
    \item We provide a novel method to use a zero-bias dense layer to visualize and analyze existing DNN models' decision boundaries. With this approach, we can diagnose the class boundaries of DNN models.
    \item We clarify the internal mechanism of abnormality detection in the zero-bias dense layer enabled DNN classifiers and provide a novel method to efficiently transfer existing DNN classifiers into DNN abnormality detectors with assured performance. 
    \item We combine our zero-bias DNN model with the Quickest Change Detection theory, and our validation on massive real signal detection demonstrates the effectiveness of our integral solution.
\end{itemize}

Our research offers a solution to accurate identification of abnormalities with assured performance, thus useful in promoting trustworthy IoT and deepening the understanding of deep neural networks. Besides, the success of the zero-bias layer enables the move from IoT to real-time control. 

The remainder of this paper is organized as follows: A literature review of related works is presented in Section~\ref{sectRW}. We formulate our problem in Section~\ref{sectPD} with the methodology presented in Section~\ref{sectMM}. Performance evaluation is presented in Section~\ref{sectEED} with conclusions in Section~\ref{sectCC}.

\section{Related works}
\label{sectRW}
Abnormality detection plays an increasingly important role in safety-critical and latency-constrained IoT, e.g., the aviation communication system in this research. Especially in the era of Big Data, multisource heterogeneous data are generated timely in huge volumes. Therefore, quick and reliable identification and detection of abnormal events and abnormalities are increasingly discussed. This section covers the state-of-art from three perspectives:

\subsection{Abnormality detection in deep neural networks}
A critical problem for learning based device identification is that classifiers only recognize pretrained data but can not deal with novel data presented during training. One intuitive alleviation is to remove the Softmax function. In \cite{wong2018clustering}, the authors first trained a CNN model with a Softmax output on known data. They then remove the Softmax function and turn the neural network into a nonlinear feature extractor. Finally, they use the DBSCAN algorithm to perform cluster analysis on the remapped features and show that the method has the potential of detecting a limited number of novel classes. 

From the perspective of Artificial Intelligence, this issue is categorized as the Open Set Recognition \cite{scheirer2012toward,bendale2016towards} problem. In \cite{roy2019rfal}, the authors use the Generative Adversarial Network (GAN) to generate highly realistic fake data. Then they exploit the discriminator network to distinguish whether an input is from an abnormal source. In \cite{shi2019deep}, the authors provide two methods to deal with abnormalities: i) Reuse trained convolutional layers to transform inputs to feature vectors, and then use Mahalanobis distance to judge the outliers. ii) Reuse the pretrained convolutional layers to transform signals to feature vectors, and then perform k-means (k = 2) clustering to discover the groups of outliers. 

\subsection{Quickest event detection}
Real-time event detection is a critical function in safety-critical IoT. From the perspective of input data, we may categorize them into single-shot and sequential detection paradigms. In single-shot detection \cite{liu2020zero}, event detections are performed per observation, and the past data will not be retained for future use. In contrast, the sequential detection paradigm allows accumulating information from past observations \cite{perera2017efficient}. 

From the perspective of the stochastic process, a Cyber-Physical System in different states can be described by distributions with measurable statistical properties \cite{lai2008quickest}. Therefore, transitions within states cause the change of those properties. The quickest detection aims to detect the change as quickly as possible, subject to false alarm constraints \cite{poor2008quickest}. The process is essentially an optimization problem. Considering whether prior observations are independent of an abnormal event's appearance, the optimization scheme can be defined in different forms as reviewed in \cite{johnson2017detecting}.

We can also categorize the quickest event detection methods into two branches: a) detecting events with known postchange distributions. b) detecting events with unknown postchange distributions. Generally, detecting known events is faster with CUSUM algorithm can be applied directly \cite{basseville1993detection}. A postchange distribution may not be known in some scenarios in advance, and nonparametric strategies have to be used and bring higher latency.

Quickest detection provides a performance-assured solution to detect change points (related to events) in sequential data. However, the selection of statistic metrics still depends on trial-and-error. We focus on real-time sequential detection of events, especially on integrating the quickest detection theory with deep learning to provide an automated and performance-assured solution to latency-constrained CPS.

\section{Problem definition}
\label{sectPD}
\begin{figure}[t]
\centering
\includegraphics[width=\linewidth]{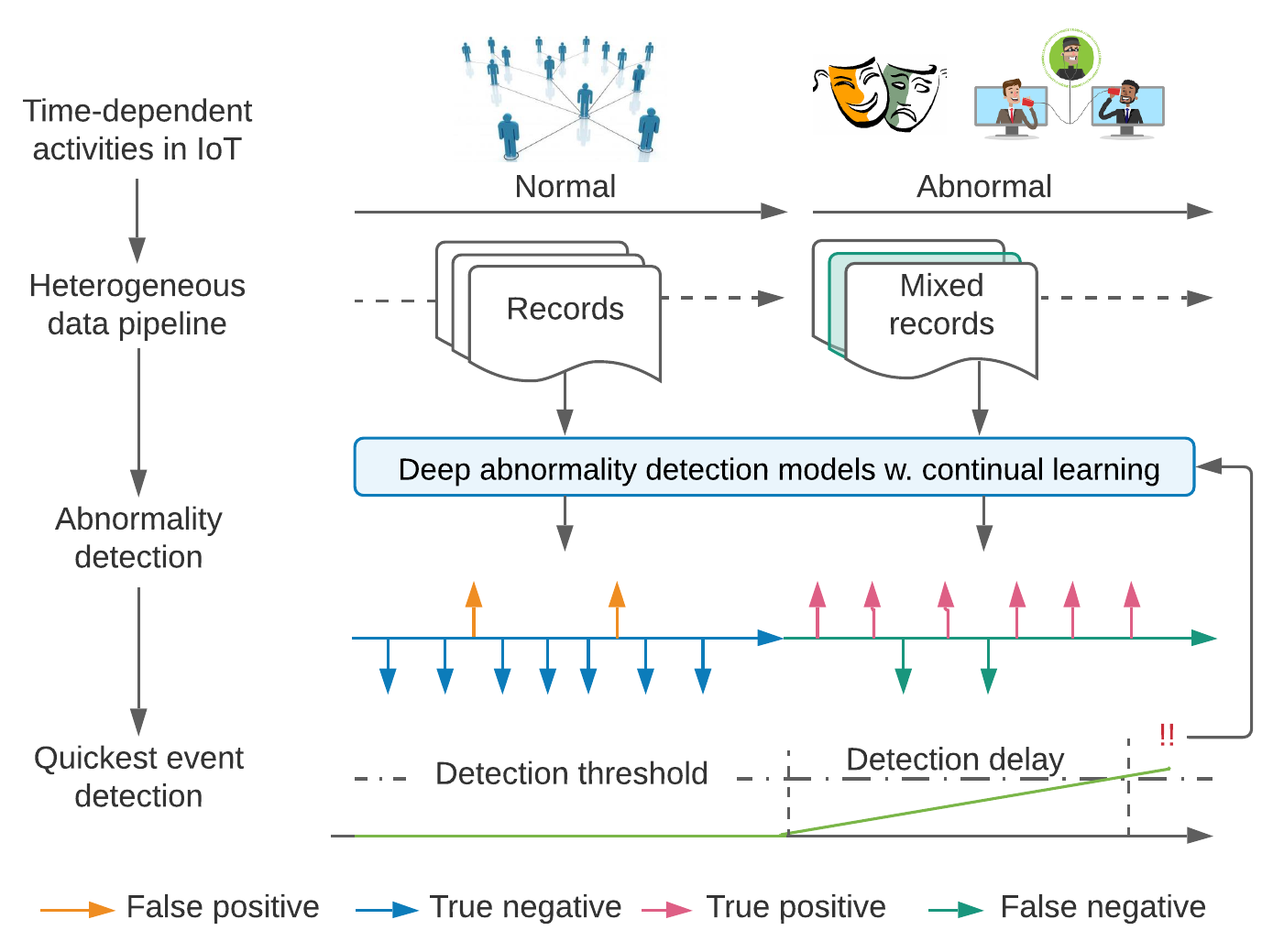}
\caption{System model of zero-bias deep learning enabled quick and reliable abnormality detection in IoT.}
\label{figSystemModel}
\end{figure}

The system model of our proposed framework is depicted in Figure~\ref{figSystemModel}. We aim to use deep learning models to process heterogeneous data from IoT and spot the abnormal data. We then use the quickest event detection algorithm to detect ongoing abnormal events with minimum latency.

In IoT systems, states are highly correlated with time-dependent events, e.g., abnormal events or normal operations. \textit{We define that abnormalities are suspicious data caused by abnormal events}. Intuitively, abnormalities could trigger variation of specific indication metrics. Analyzing the drift or variations of these metrics, abnormal events can be detected sequentially. Therefore, we can convert the real-time abnormality detection problem into an online sequential event detection scheme, in which a surveillance oracle can sequentially collect its target system's state signals or heterogeneous data denoted as:
\begin{align}
\label{eqObservation}
    X = \{\boldsymbol{X_1}, \boldsymbol{X_2}, \dots \boldsymbol{X}_j \dots \boldsymbol{X}_{j+m} \dots\}
\end{align}
where $\boldsymbol{X_j}$ denotes a state variable or record in vector form, an abnormal event appears at $j$ and disappear at $j+m$. Real-time abnormal event detection requires triggering an alarm before $j+m$ with minimum assured latency. 

Well-known methods are provided in the Quickest Change Detection (QCD) theory. For example, in the Cumulative Sum (CUSUM) Control Chart algorithm, a likelihood ratio test is employed to sequentially process the observed data at each timestamp $k$, denoted as:
\begin{align}
    g(k) = ln(\dfrac{P_1(\boldsymbol{X}_k)}{P_0(\boldsymbol{X}_k)})
\end{align}
Where $g(k)$ is a sufficiency metric, $P_0(\cdot)$, $P_1(\cdot)$ denotes the probabilistic density functions of abnormal and abnormal states, respectively. A constrained cumulative sum of sufficiency metrics is used as an indicator, denoted as:
\begin{align}
    S(k) = max(0,S(k-1) + g(k))
\end{align}
An alarm will be sent once $S(k)$ is greater than a predefined threshold, $h$. The CUSUM algorithm has been proved to provide the lowest worst-case detection latency at specific false alarm intervals. However, CUSUM-style quickest detection algorithms can hardly handle high-dimension data, where $P_0(\cdot)$ and $P_1(\cdot)$ are difficult to obtain. Even though some works use DNNs to derive the sufficiency metric $g(k)$ from high dimension data, the DNNs' uncertain responses when encountering abnormalities make performance assurance a theoretic challenge. To enable deep learning for quick and reliable abnormality detection, the following efforts are needed:
\begin{itemize}
    \item We need a DNN driven abnormality detection model to process complex data and provide theoretically assured performance. If possible, the deep abnormality detection model should be derived without a large overhead.
    \item We need to develop an efficient method to jointly apply performance-assured DNN and quickest event detection to provide theoretically guaranteed performance in detecting abnormal events.
\end{itemize}

\section{Proposed Framework}
This section will first introduce the zero-bias DNN and its application in the explainability of DNN models. We then provide a method to convert zero-bias DNN into a performance assured abnormality detector efficiently. Finally, we provide our method to integrate zero-bias DNN with the quickest detection algorithms.
\label{sectMM}
\subsection{DNNs with zero-bias dense layers}
\label{subsectDNNZb}
This subsection provides an extended analysis of the zero-bias dense layer's characteristics, which serves theoretic backgrounds. Additionally, we show that the decision boundaries of DNNs with zero-bias dense layers can be visualized conveniently. Hereon, we will use the term DNNs with zero-bias dense and zero-bias DNN alternatively.

We have discovered that the last dense layers of a DNN classifier perform the nearest neighbor matching with biases and preferabilities using cosine similarity in \cite{liu2020zero}. We then show that DNN classifiers' accuracy will not be impaired if we replace their last dense layers with our zero-bias dense layers, in which biases and preferabilities are eliminated. We can denote the mechanism of the zero-bias dense layer as:
\begin{align}
\label{eqZeroBiasDenseLayer}
    \boldsymbol{Y}_0(\boldsymbol{X}) &= \boldsymbol{W}_0\boldsymbol{X} + \boldsymbol{b}\\\notag
    L(\boldsymbol{X}) &= cos(\boldsymbol{Y}_0, \boldsymbol{W}_1)
\end{align}

Where $\boldsymbol{X}$ is the output of the prior layer, a.k.a., feature vectors. $\boldsymbol{W}_1$ is a matrix to store fingerprints of different classes. $\boldsymbol{X}$ is an $N_0$ by $q$ matrix, where $N_0$ denotes the number of features while $q$ denotes the batch size. $\boldsymbol{W}_0$ is an $N_1$ by $N_0$ matrix where $N_1$ denotes the number of new feature dimensions. $\boldsymbol{W}_0\boldsymbol{X} + \boldsymbol{b}$ performs linear dimension reduction as long as $N_1 < N_0$. Finally, $W_1$ is a $C$ by $N_1$ matrix in which $C$ denotes the number of classes, Please be noted that in $W_1$, each row represents a fingerprint of corresponding class whilst in $\boldsymbol{Y}_0$ each column represents a feature vector within a batch. Therefore, the cosine similarity can be implemented by:
\begin{align}
\label{eqCosine}
    cos(\boldsymbol{Y}_0, \boldsymbol{W}_1) =\boldsymbol{RU(W_1)} \times \boldsymbol{CU(Y_0)}
\end{align}

Where $\boldsymbol{RU(\cdot)}$ and $\boldsymbol{CU(\cdot)}$ denote deriving column-wise and row-wise unified (vectors' magnitudes normalized to one) vectors of the input matrix, respectively. Our prior results \cite{liu2020zero, liu2020deep} also prove that zero-bias DNN can be trained using common loss functions (e.g., binary crossentropy, MSE, and etc.) and back-propagation mechanisms.

\begin{figure}[h]
\centering
\includegraphics[width=0.5\linewidth]{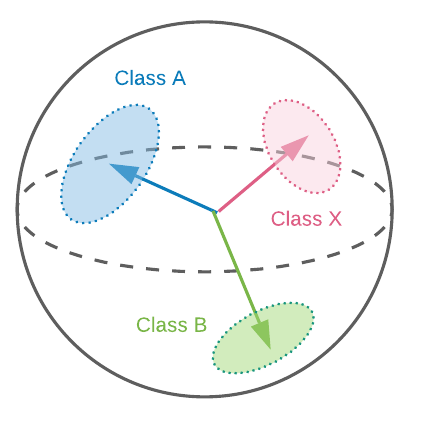}
\caption{3D unit hyperspherical surface in zero-bias DNNs.}
\label{figHypersphere}
\end{figure}

\begin{figure}[h]
\centering  
\subfloat[85\% accuracy]
{%
    \includegraphics[width=0.95\linewidth]{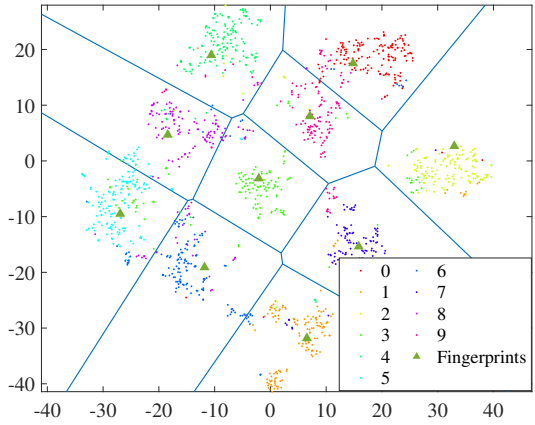}
    \label{figBadlyTrained}
}\\
\subfloat[97\% accuracy]
{%
    \includegraphics[width=0.95\linewidth]{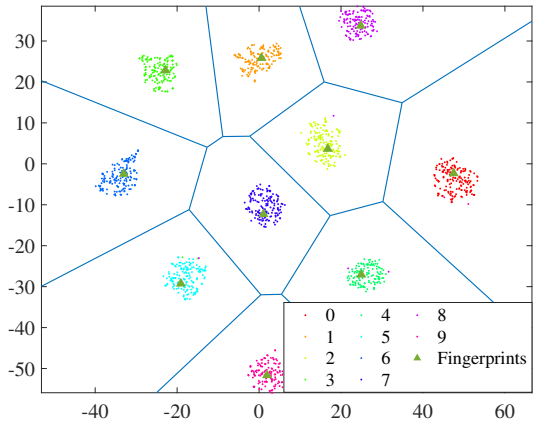}
    \label{figWellTrained}
}
\caption{The Voronoi diagrams of dimension-reduced fingerprints and validation set at different training stages. Fingerprints and feature vectors are projected to a 2D space using \textit{t-SNE} algorithm \cite{maaten2008visualizing}.  }
\label{figVoronoiTraining}
\end{figure}

The cosine similarity in Equation \ref{eqCosine} represents the similarity matching of fingerprints and feature vectors on an $N_1$-D unit hyperspherical surface. A 3-D example is depicted in Figure~\ref{figHypersphere}. Fingerprints divide the unit hyperspherical surface into several subregions, and we can reduce the dimension of fingerprints and use Voronoi Diagram \cite{erwig2000graph} to visualize their decision boundaries as in Figure~\ref{figVoronoiTraining}. The decision boundaries of a specific fingerprint denote the boundaries of a class. Hereon, we will use the terms class boundaries and fingerprint's boundaries alternatively.

Please be noted that, even if we eliminate the magnitude and bias constant for each fingerprint, we do not need to worry about the capacity of zero-bias DNN models in distinguishing different classes. A numerical example for quantifying the maximum theoretic number of classes that a zero-bias DNN will reliably learn is given in Figure~\ref{figMaximumNumClasses}. .

For example, in the DNN enabled MNIST handwritten digit recognition \cite{matlabMinst}, the network's last dense layer is replaced by a zero-bias dense layer with $N_1 = 10$. The Voronoi diagrams at two stages (85\% and 97\% accuracies) using fingerprints in the zero-bias dense layer and feature vectors from the validation set are depicted in Figure \ref{figVoronoiTraining} with solid blue lines representing the decision boundaries of topologically adjacent classes. Please be noted that the Voronoi diagram can not be applied directly to DNN models without adapting zero-bias dense layers.

From the observation, we conclude that during training, a DNN with a zero-bias dense layer learns to project input data from identical classes closer to the corresponding fingerprints and separate data from different classes far away. The feature extractors in prior layers and fingerprints in the zero-bias dense layer are optimized simultaneously. In DNNs with zero-bias dense layer, classification errors result from two perspectives:
\begin{itemize}
    \item If fingerprints are not distantly separated, data from the corresponding classes are highly possible to get confused. This fact is verified in our previous work in \cite{liu2020zero}.
    \item The prior layers are poorly trained and the feature vectors are sparsely projected as depicted in Figure~\ref{figBadlyTrained}.
\end{itemize}

Please be noted that zero-bias dense layer can be easily adapted to existing DNN models via transfer learning.

\subsection{Abnormality detection in DNN with zero-bias dense layer}
The effectiveness of zero-bias DNN for nonsequential abnormality detection has been demonstrated in our prior results \cite{liu2020zero, liu2020deep}, briefly, it is significant better than regular DNN and comparable to one-class SVM \cite{choi2009least}. In this section, we deepen our previous research and present a solution to convert zero-bias DNN models into abnormality detectors with predictable and assurable performance.

\subsubsection{Deriving abnormality detector from existing DNN classifiers}
\begin{figure}[h]
\centering  
\subfloat[Abnormalities from unknown classes]
{%
    \includegraphics[width=0.4\linewidth]{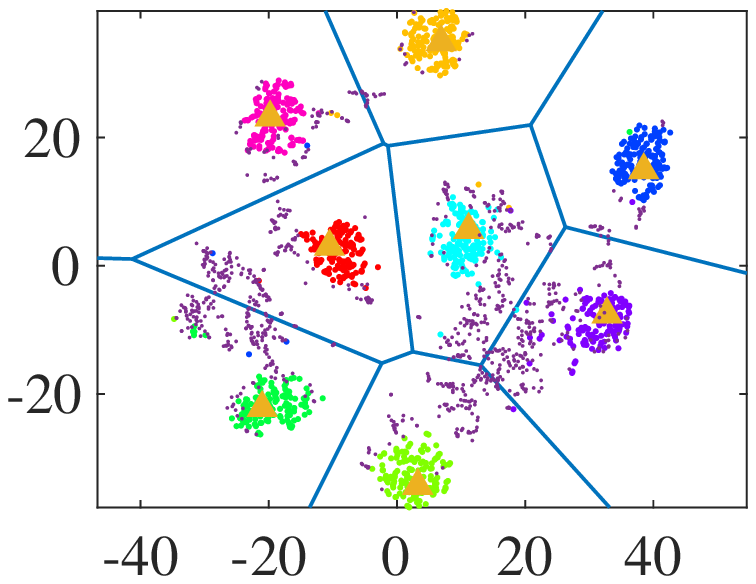}
    \label{figDistribAbnormalities}
}\hfil
\subfloat[Abnormalities from random noise $\sim N(0,2)$]
{%
    \includegraphics[width=0.4\linewidth]{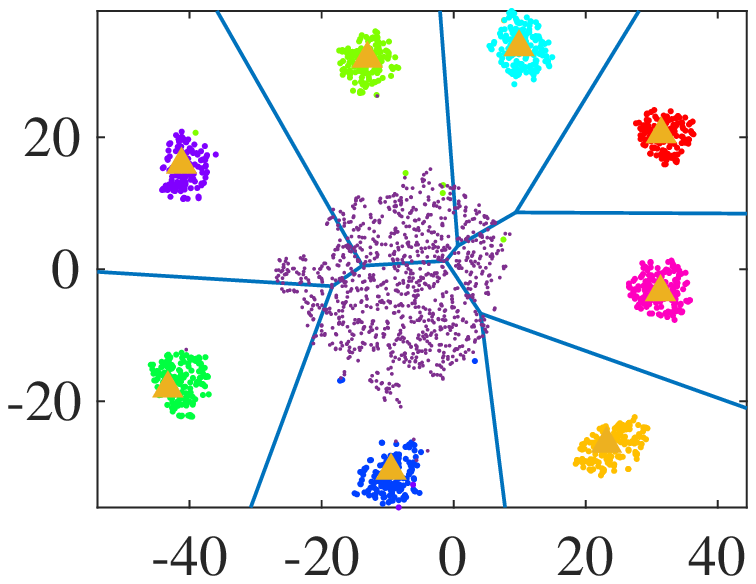}
    \label{figDistributionRandomNum}
}
\caption{Voronoi diagrams of dimension-reduced fingerprints, validation set and abnormal data. Data are projected to a 2D space using \textit{t-SNE} algorithm \cite{maaten2008visualizing}.  }
\label{figVoronoiAbnormalities}
\end{figure}

In Figure~\ref{figWellTrained}, feature vectors from known classes are closely projected to the vicinity of the corresponding fingerprints. As a result of cosine similarity matching, we come to our first remark:
\begin{remark}
\label{rmFingerprintVicinity}
Abnormal data from an unknown novel class are less likely to be projected into any existing classes' close vicinity as there is no specific fingerprint for these data.
\end{remark}
We use the MNIST example to demonstrate Remark~\ref{rmFingerprintVicinity} with the results in Figure~\ref{figVoronoiAbnormalities}. We train the zero-bias DNN to recognize handwritten digits from 1 to 8 and use digits 9 and 0 as abnormal data. The projection feature vectors of known and abnormal data and fingerprints are depicted in Figure~\ref{figDistribAbnormalities}. We then replace the abnormal data with pure Gaussian random noise ($N(0,2)$) and repeat the experiment. Results are depicted in Figure~\ref{figDistributionRandomNum}.

\begin{figure}[h]
\centering
\includegraphics[width=0.8\linewidth]{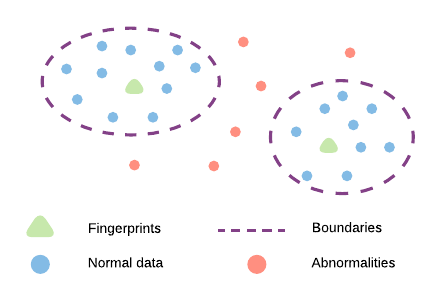}
\caption{Relation of normal and abnormal data.}
\label{figClassModeling}
\end{figure}

It can be noticed that abnormalities from an unknown class of the same domain could be even more difficult to detect than pure random noise. Although Figure~\ref{figDistribAbnormalities} has shown that abnormalities from unknown classes are more sparsely distributed in the unit hyperspherical surface, Remark~\ref{rmFingerprintVicinity} still holds. Therefore, we can derive a basic principle (depicted in Figure~\ref{figClassModeling}. ) to convert a zero-bias dense layer enabled DNN classifier into an abnormality detector:
\begin{remark}
We can model the spatial distribution and boundaries of normal data in the hyperspherical surface. Then the incoming feature vectors that are out of normal data boundaries are regarded as abnormalities.
\end{remark}

For a given DNN model with zero-bias dense layer, we model the boundaries of different classes as follows:
\begin{enumerate}[\textbf{Step} 1:]
\item The training set is utilized to learn the boundaries of known classes while the validation set will be mixed with abnormal data ($\boldsymbol{A_0}$) to measure the performance of converted abnormality detector.
\item We pass accurately classified data of each known class in $\boldsymbol{C_1}$, denoted as $\boldsymbol{KX}_i$, through the DNN model and obtain the compressed feature vectors before fingerprint matching, denoted as:
\begin{align}
    \boldsymbol{Y}_0[F_{n-1}(\boldsymbol{KX}_i)] = \boldsymbol{W}_0 F_{n-1}(\boldsymbol{KX}_i) + \boldsymbol{b}
\end{align}
Where $\boldsymbol{W}_0$ and $\boldsymbol{b}$ are defined in Equation~\ref{eqZeroBiasDenseLayer}, $F(\cdot)_{n-1}$ denotes all network layers before the fingerprint matching. $\boldsymbol{Y}_0[F_{n-1}(\boldsymbol{KX}_i)]$ denotes feature vectors of accurately classified data in $\boldsymbol{KX}_i$.
\item Calculate the centroid $\boldsymbol{c}_0^i$ and covariance matrix ($\boldsymbol{P_i}$) of $\boldsymbol{KX}_i$ as:
\begin{align}
\boldsymbol{c}_0^i &= mean( \boldsymbol{Y}_0[F_{n-1}(\boldsymbol{KX}_i)])\\\notag
\boldsymbol{P_i} &= cov( \boldsymbol{Y}_0[F_{n-1}(\boldsymbol{KX}_i)], \boldsymbol{Y}_0[F_{n-1}(\boldsymbol{KX}_i)])
\end{align}
\item Calculate the Mahalanobis distances \cite{de2000mahalanobis} from the class centroid $\boldsymbol{c}_0^i$ to all accurately classified feature vectors. Then we use the maximum value as a cut-off distance $CO_i$ of class $\boldsymbol{KX}_i$:
\begin{align}
    CO_i = max~\boldsymbol{D_m}[\boldsymbol{Y}_0[F_{n-1}(\boldsymbol{KX}_i)],\boldsymbol{c}_0^i]
\end{align}
Where $\boldsymbol{D_m}(\cdot, \boldsymbol{c}_0^i)$ denotes the feature vectors' Mahalanobis distances to $\boldsymbol{c}_0^i$.
\item Abnormality detection using cut-off boundaries on input data $\boldsymbol{X})$ is formally defined as:
\begin{align}
\hspace*{\dimexpr-\leftmargini-\leftmarginii-\leftmarginiii}
    D(\boldsymbol{X}) = 
    \begin{cases}
     1~~\exists~i,~ \boldsymbol{D_m}[\boldsymbol{Y}_0[F_{n-1}(\boldsymbol{X})],\boldsymbol{c}_0^i] \leq CO_i\\
    0~~\text{Otherwise}
    \end{cases}
\end{align}
\end{enumerate}

These steps convert zero-bias DNNs into abnormality detectors with binary outputs. Herein, we will use the term zero-bias abnormality detector alternatively. In essence, we construct statistical models for each class to describe the distribution of corresponding normal data and a hard cut-off distance to form its boundary (denoted as dashed purple lines in Figure~\ref{figClassModeling}). Please be noted that other distance functions or modeling methods such as the Local Outlier Factor \cite{breunig2000lof} can also be applied. 

\begin{figure}[h]
\centering  
\subfloat[Coverage ratio per class]
{%
    \includegraphics[width=0.85\linewidth]{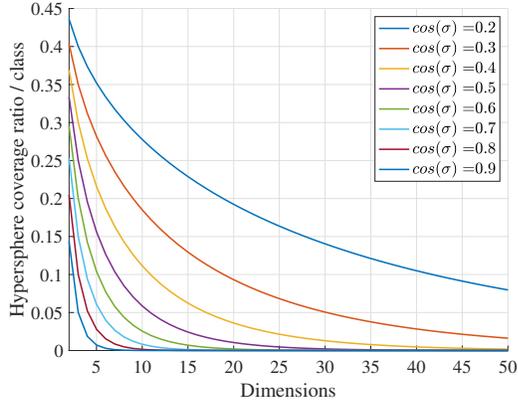}
    \label{figHyperCoverage}
}\\
\subfloat[Maximum of distinguishable classes]
{%
    \includegraphics[width=0.85\linewidth]{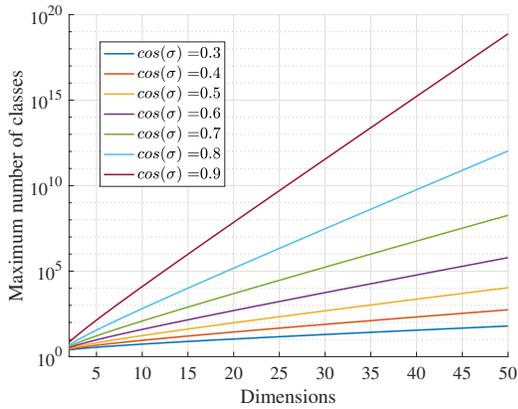}
    \label{figMaximumNumClasses}
}
\caption{The coverage ration per class and maximum number of distinguishable class in zero-bias DNN.}
\label{figZbCapacity}
\end{figure}

Suppose that each fingerprint governs a non-overlapped subregion with a maximum acceptable deviation angle, $\sigma$, for normal data. This subregion is named $\sigma$-cap, we can analytically evaluate the area of each $\sigma$-cap, and its occupied area ratio, $r_0(m)$, as:
\begin{align}
    \notag A_c(m) &= \dfrac{1}{2}A_u(m)r^{n-1}I_{(2rh-h^2)/r^2}\left(\dfrac{m-1}{2},\dfrac{1}{2}\right)\\
    \notag A_u(m) &= \dfrac{2\pi^{n/2}}{\Gamma(n/2)}\\
    r_0(m) &= \dfrac{A_c(m)}{A_u(m)} = \dfrac{1}{2}I_{(2h-h^2)}\left(\dfrac{m-1}{2},\dfrac{1}{2}\right)
\end{align}
Where $A_c(m)$ is the area of a $m$-D $\sigma$-cap. $A_u(m)$ is the surface area of the $m$-D unit hypersphere. Additionally, we have $r=1$ and $h=r-rcos(\sigma)$. $I$ and $\Gamma$ are the regularized incomplete beta function and the gamma function, respectively. A numerical result is given in Figure~\ref{figZbCapacity}. As depicted, both a smaller $\sigma$ and larger number of feature dimensions ($N_1$) increase the capacity and interclass distinguishability of the zero-bias DNN. Moreover, even if we eliminate some information of feature vectors in zero-bias DNNs, we do not have to worry much about their learning capacity and remaining space for abnormalities, as long as feature dimension $N_1$ is large.

\subsubsection{Theoretic performance Analysis}\label{subsubsectTheoreticAD}
We introduce the hard cut-off distances of fingerprints. Therefore, a binary abnormality detector converted from zero-bias DNN becomes a binary classifier. We derive two important properties of this type of zero-bias abnormality detector regarding false positive and false negative rates.  

\begin{figure}[h]
\centering
\includegraphics[width=0.8\linewidth]{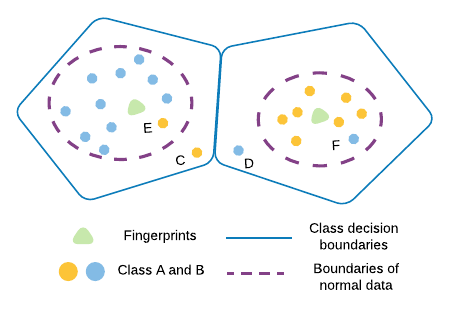}
\caption{Classification errors in zero-bias DNN, feature vector C, D, E and F are erroneously projected into governing regions of wrong fingerprints.}
\label{figClassificationError}
\end{figure}

The accuracy of zero-bias DNN models on known classes can be obtained after training. As discussed earlier in Section \ref{subsectDNNZb}, the classification errors are caused by inaccurate projections. From the perspective of decision boundary and class boundary, the scenarios that lead to classification error are depicted in Figure~\ref{figClassificationError}. As depicted, the feature vectors C and D are projected into the wrong class boundaries but out of the boundaries of normal data. Meanwhile, E and F are projected into the normal data boundaries of wrong fingerprints.

Suppose that E and F in Figure~\ref{figClassificationError} are moved out of the normal data boundaries. The false-positive rate of abnormality detection reaches its upper bound and equals the classification error $\alpha$. Furthermore, if C and D are moved into normal data boundaries, the false positive rate equals zero. Therefore, the range of false-positive rate of zero-bias abnormality detector is actually determined:
\begin{remark}[Range of the false positive rate]
\label{rmFPRAcc}
Suppose that the classification error of the zero-bias DNN is $\alpha$, as long as our statistical model can closely follow the boundary of normal data, the false positive rate of converted abnormality detector is less than or equals to $\alpha$. Denoted as:
\begin{align}
    FPR \leq \alpha
\end{align}
\end{remark}

Suppose that in a regular case, the feature vectors of abnormalities are mixed with normal data and uniformly distributed on the surface of the unit hypersphere, in this case, the maximum false negative rate is reached.
\begin{remark}[Range of false negative (true positive) rates]
\label{rmTPFNR}
The upper bound of the false negative rate under uniformly distributed abnormalities, equals to ratio of the occupied regions' area of normal data divided by the total surface area of the unit hypersphere, denoted as:
\begin{align}
    \notag &RU_{FNR} = \dfrac{\sum_{i=1}^{N_c} S^i_{hsp}(N_1)}{A_{hsp}(N_1)}\\
    \text{With}~&FNR\leq RU_{FNR},~TPR \geq 1-RU_{FNR}
\end{align}
Where $N_c$ is the number of known classes, $N_1$ and $A_{hsp}(N_1)$ are the dimension and surface area of the unit hypersphere, respectively. $S^i_{hsp}(N_1)$ is the surface area of normal data of the $i$th class. 
\end{remark}

Analytically calculating $S^i_{hsp}(N_1)$ is difficult since the shapes of these occupied subregions are unknown. Therefore, we use Monte Carlo method to estimate $RU_{FNR}$ directly. Corresponding pseudo code is presented in Algorithm~\ref{algMonteCarloRu}, specifically, we generate $M$ random points uniformly distributed on the surface of the unit hypersphere and count the ratio of points that are captured by the normal data boundaries of fingerprints. The captured rate directly indicates the value of $RU_{FNR}$. Empirically, we set $M = 20,000$.
\begin{algorithm}
\caption{Estimating $RU_{FNR}$}
\label{algMonteCarloRu}
\begin{algorithmic}[1]
\Function{$RU_{FNR}$}{$N_1,N_c,M,List[\boldsymbol{CO}],List[\boldsymbol{c}^i_0]$}
   \State $HX\gets UniformHypersphereRand(N_1,M)$ \Comment{Please refer to \cite{hypersphere}}
   \State $chx \gets 0$
   \For {$k\gets 1 \dots M$}
        \For {$i \gets 1 \dots N_c$}
            \If {$\boldsymbol{D_m}[HX_k,\boldsymbol{c}^i_0]\leq \boldsymbol{CO}_i$}
            \State $chx \gets chx + 1$
            \EndIf 
        \EndFor
   \EndFor
   \State \textbf{return} $\dfrac{chx}{M}$
\EndFunction
\end{algorithmic}
\end{algorithm}

\subsection{Zero-bias DNN for quickest abnormal event detection}
\subsubsection{Sequential formalization and detectability}
\label{sectBernoulliFormalization}
Given the theoretic analysis of zero-bias abnormality detector in section~\ref{subsubsectTheoreticAD}, we can model the response of zero-bias DNNs as switching between two probability distributions before and after the appearance of an abnormal event, namely $P_0$ and $P_1$, respectively. Since we have converted the zero-bias DNN into a binary abnormality detector, we can formulate $P_0$ and $P_1$ into two Bernoulli Distributions \cite{weisstein2002bernoulli}:
\begin{align}
    \notag P_0(I_k) &= FPR^I_k(1-FPR)^{1-I_k}\\
    \notag P_1(I_k) &= (1-FNR)^I_kFNR^{1-I_k}\\
    &=(TPR)^I_k(1-TPR)^{1-I_k}
\end{align}
Where $I \in \{0,1\}$ is the binary output of the abnormality detector with $I_k = D(\boldsymbol{X_k})$. $FPR$ can be deried on existing data, and the range of $FNR$ and $TPR$ from section~\ref{subsubsectTheoreticAD}. As long as the $P_0$ and $P_1$ are different, the abnormal event causing drifts from $P_0$ to $P_1$ can be sequentially detected. We have the following determinant under regular scenarios:
\begin{remark}[Sequential detectability]
\label{rmSeqDetectability}
Abnormal events are assured to be sequentially detectable if the binary zero-bias detector's true-positive rate ($TPR$) lower bound ($1-RU_{FNR}$) are greater than the false-positive rate ($FPR$) upper bound ($\alpha$).
\end{remark}

\begin{figure}[h]
\centering
\includegraphics[width=0.9\linewidth]{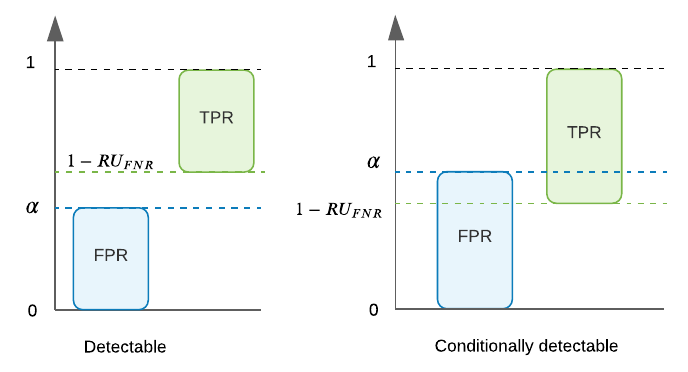}
\caption{Range of true-positive and false-positive rates.}
\label{figTPRFPRRange}
\end{figure}

Remark~\ref{rmTPFNR} shows that the true positive and false negative rates are within different ranges, if $1-RU_{FNR} \leq \alpha$, the two variables' spanning ranges are partially overlapped (depicted in Figure~\ref{figTPRFPRRange}) and we may encounter an extreme case: $TPR = FPR$. Therefore, the abnormal event is only conditionally detectable. 

\subsubsection{Quickest detection algorithm}
With Remark~\ref{rmSeqDetectability}, we can use Quickest Change Detection algorithm to detect the appearance of an abnormal event with the lowest latency at a given false alarm run length. We will present both the Bernoulli Generalized Likelihood Ratio Chart and its approximation, the multiple Bernoulli CUSUM Chart, respectively. Compared with the existing nonparametric solutions, we discretize the continuous probabilistic function space and transform the problem into a parametric sequential hypothesis testing problem.

Using Bernoulli Generalized Likelihood Ratio (GLR) Chart \cite{huang2013generalized} to sequentially detect abnormal events. We have:
\begin{align}
    \notag R_k &= \max_{0 \leq \tau \leq k-1, \beta \leq TPR \leq 1} \ln \dfrac{\prod_{i=\tau+1}^kTPR^{I_k} (1-TPR)^{1-I_k}}{\prod_{i=\tau+1}^k FPR^{I_k} (1-FPR)^{1-I_k}}\\\notag
    &=\max_{0 \leq \tau \leq k-1}(k-\tau)\ln \left [ \widehat{TPR}\cdot \dfrac{\widehat{TPR}(1-FPR)}{FPR(1-\widehat{TPR})} \right.\\
    &~~~~~~+ \left. \ln\dfrac{1-\widehat{TPR}}{1-FPR}\right]
\end{align}
Where $\widehat{TPR}\approx TPR \in [1-RU_{FNR},1)$ is the estimated true positive rate of zero-bias abnormality detector and $\tau$ is the estimated time when an abnormal event happens. $\widehat{TPR}$ is dynamicaly estimated as follows:
\begin{align}
    \widehat{TPR} &= min\left \{B_1 ,max \left[1-RU_{FNR},\dfrac{\sum_{i=\tau + 1}^k}{k-\tau}I_k \right]\right \}
\end{align}
Where $B_1 = 1-\varepsilon$ is the maximum possible value of $TPR$ and $\varepsilon$ is a tiny positive number to assure $\widehat{TPR} < 1$. An alarm is triggered if $R_k > h_{GLR}$ and $h_{GLR}$ is a pre-defined threshold. $h_{GLR}$ can be chosen as suggested in: \cite{huang2013generalized}:
\begin{align}
    h_{GLR} = log_{10} (ARL\cdot FPR)
\end{align}
Where $ARL$ is the average run length between false alarms.

Theoretically, we have to store a long sequence ($0 \leq \tau \leq k-1$) of previous abnormality detection results to detect an abnormal event. Fortunately, we can use a sliding window to store relevant data and reduce the computational complexity. In \cite{lai1998information} and \cite{huang2013generalized}, it is shown that a GLR chart with a window is asymptotically optimal if the window size $m$ is sufficiently large.

It is also numerically verifiable that the detection latency of Bernoulli GLR charts can be closely approximated with a countable set of Bernoulli CUSUM Charts, where the identical detection threshold $h_{CUSUM}$ is shared among them and $h_{CUSUM} = h_{GLR}$ \cite{huang2012binomial, huang2013generalized}. The approximated range of $TPR$ is covered by each CUSUM chart is:
\begin{align}
    \widehat{TPR}_i = 1-RU_{FNR} + \dfrac{TPR_{max} \cdot i^2}{U^2}
\end{align}
Where $U$ is the total number of CUSUM charts, in which greater than 100 is recommended, $i$ denotes the index of each chart. $TPR_{max}$ is the max possible value of the true positive rate that is less than 1. $1-RU_{FNR}$ denotes the lower bound of the true positive rate. Therefore, given an average run length between false alarms, $ARL$, we have the worst case average detection delay as:
\begin{align}
    \Bar{T}_{GLR} = \Bar{\boldsymbol{T}}_{CUSUM}  \sim \dfrac{h_{CUSUM}}{I(P_1,P_0)}
\end{align}
Please be noted that we use the characteristic of multiple Bernoulli CUSUM charts to demonstrate the properties of detection delay.


\section{Performance Evaluation}
\label{sectEED}
In this section, we evaluate the performance of the proposed framework in two folds, we first use a massive real-world signal dataset to test the proposed method for visualizing the class decision boundaries and feature vectors of a deep signal identification network \cite{yongxin_liu_2020_4248678}. Then we use the proposed method to convert the identity recognition DNN into an abnormality detection model and evaluate its performance on sequential abnormal event detection.
\subsection{Dataset and application scheme}
\begin{figure}[h]
\centering
\includegraphics[width=0.9\linewidth]{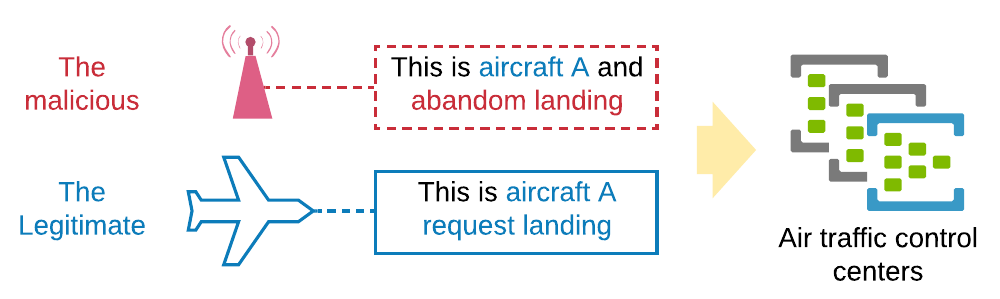}
\caption{Identity spoofing in aviation communication systems.}
\label{figAttackATC}
\end{figure}

Our dataset is available in \cite{gt9v-kz32-20}, we use the wide-spreading signals from Automatic Dependent Surveillance-Broadcast (ADS-B) system \cite{riddle1090}, which provides a great variety of signals from commercial aircraft's signal transponders with labels. Specifically, each licensed aircraft use 1090MHz transponders to broadcast their geo-coordinates, velocity, altitude, heading, as well as its unique identifier to the Air Traffic Control (ATC) center. The integrity and trustworthiness of ADS-B signals are critical to aviation safety. However, the ADS-B system does not contain cryptographical identity verification mechanisms and thus is vulnerable to identity spoofing attacks (depicted in Figure~\ref{figAttackATC}). Our previous work shows that the responses of the zero-bias DNN on known (learned) and unknown transponders' signals are different. Therefore, we can use the methodology provided in this paper to design a performance-assured quick abnormality detector to detect signals from the malicious masquerading legitimate aircraft.

From the perspective of Deep Learning, the input is the raw signal from a Software Defined Radio Receiver (USRP B210) and the label is the identify of the signal source (aircraft). As in our prior work \cite{liu2020zero, liu2020deep}, we take the first 1024 samples from the signal of each intercepted message. And convert the 1024 samples into a 32 by 32 by 3 tensor, which incorporate pseudo noise, magnitude-frequency domain, and phase-frequency domain information, respectively. 

\begin{figure}[h]
\centering
\includegraphics[width=0.95\linewidth]{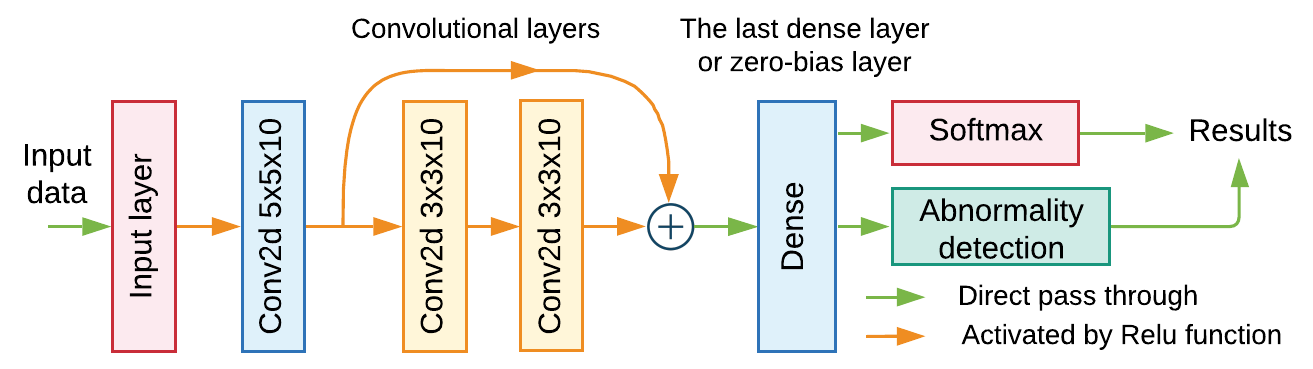}
\caption{Deep neural network architecture.}
\label{figLearningArch}
\end{figure}

\subsection{Decision boundaries and feature vectors in real-world zero-bias DNN}
\begin{figure}[h]
\centering  
\subfloat[Accuracy 93.7\% under polluted signals]
{%
    \includegraphics[width=0.85\linewidth]{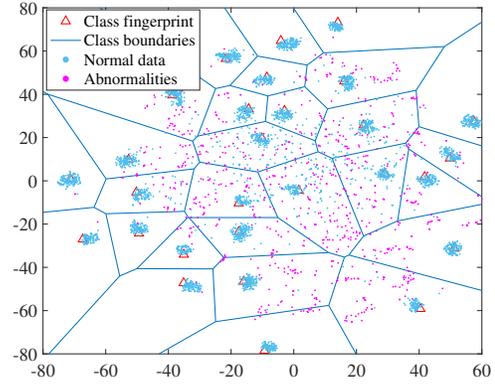}
    \label{figAbnormalInRealWorld}
}\\
\subfloat[Accuracy 99.9\% under filtered signals]
{%
    \includegraphics[width=0.85\linewidth]{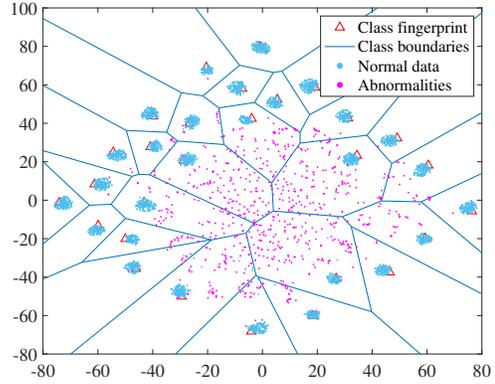}
    \label{figAbnormalInRealWorld100}
}
\caption{Voronoi diagram of dimension-reduced fingerprints and validation set two different scenarios. Fingerprints and feature vectors are projected to a 2D space using \textit{t-SNE} algorithm \cite{maaten2008visualizing}.  }
\label{figVoronoiTwoScenarios}
\end{figure}

The architecture of our DNN model is depicted in Figure~\ref{figLearningArch} with a description of the dataset in Table~\ref{tabDataUsage}. In Figure~\ref{figVoronoiTwoScenarios} we use two Voronoi diagrams to depict the relation of fingerprints, class boundaries, normal and abnormal data, with the DNN model trained under two scenarios: a) Input signals are polluted by abrupt spike noise due to the signal interference. b) Input signal without abrupt spikes (removed by a Gaussian filter).

\begin{table}[h]
\caption{Description of dataset}
\label{tabDataUsage}
\centering
\begin{tabular}{cl}
\toprule
Usage & \multicolumn{1}{c}{Description} \\ \midrule
Training & \begin{tabular}[c]{@{}l@{}}60\% of signals, including 28 aircraft with more than \\500 randomly selected raw records for each.\end{tabular} \\ \midrule
Validation & \begin{tabular}[c]{@{}l@{}}40\% of signals, including 28 aircraft with more than \\500 randomly selected raw records for each.\end{tabular} \\ \midrule
Normal data & \begin{tabular}[c]{@{}l@{}}Validation data that's not been used to train the\\ network.\end{tabular} \\ \midrule
Abnormal data & \begin{tabular}[c]{@{}l@{}}Signals including 236 aircraft with less than 200 raw \\signal records during the data collection period.\end{tabular} \\ \bottomrule
\end{tabular}
\end{table}

As in Figure \ref{figAbnormalInRealWorld} and \ref{figAbnormalInRealWorld100}, normal data are closely distributed within their fingerprints while abnormalities are sparsely distributed over the feature space. It is interesting to find that if the DNN model is with high accuracy, the abnormalities are less likely to appear in normal data clusters. The DNN signal identifier is with high accuracy within the two scenarios. 

According to Remark~\ref{rmFPRAcc} and \ref{rmTPFNR}, the abnormality detector's performance is predictable. In reality, the zero-bias abnormality detector trained on noisy data has a true positive and a true negative rate of 91\% and 92\%, respectively, which is closely matched with our prediction. However, the zero-bias abnormality detector trained on the filtered data has a true positive and a true negative rate of 99\% and 91\%. The true negative rate is smaller than the expected value due to the model entering the early stage of overfitting. At this stage, the training set's feature vectors can no longer provide sufficient information on the distribution of normal data. Therefore, the estimation of normal data will be misled. 

\begin{figure}[h]
\centering
\includegraphics[width=0.85\linewidth]{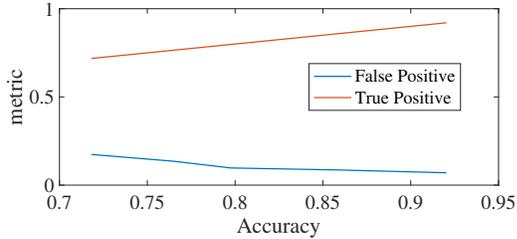}
\caption{Performance of the converted abnormality detector.}
\label{figPerformanceMetricAbn}
\end{figure}

For alleviation, we can set a threshold value of accuracy during training. We derive the cut-off distances, centroids, and covariance matrices at this point. By setting a triggering value of 96\% for the validation accuracy on the filtered data, we get a true-positive rate and a false-positive rate of 98\% and 95\%, respectively. The relation between the performance of the converted abnormality detector and the zero-bias DNN model's accuracy before conversion is given in Figure~\ref{figPerformanceMetricAbn}. As predicted, when the accuracy of zero-bias DNN gets higher, the normal data occupies a smaller amount of area on the unit hypersphere surface, and thus produce higher True Positive rates. Meanwhile, lower False Positive rates are achieved when the zero-bias DNN has higher classification accuracy before conversion as predicted.

\begin{figure}[h]
\centering  
\subfloat[]
{%
    \includegraphics[width=0.85\linewidth]{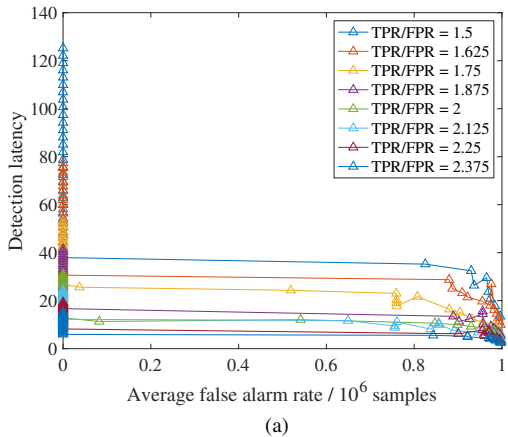}
    \label{figLatencyVsARL}
}\\
\subfloat[]
{%
    \includegraphics[width=0.85\linewidth]{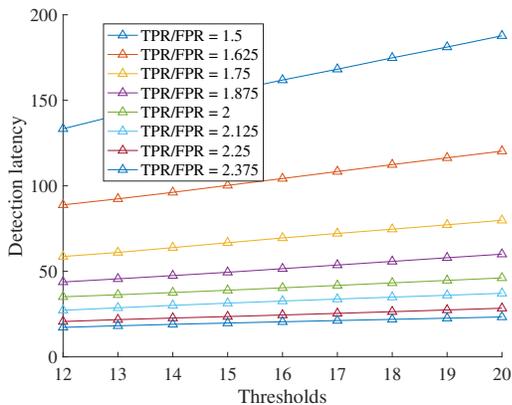}
    \label{figLatencyVsThresholds}
}\\
\caption{Abnormal event detection latencies.}
\label{figSequentialLatency}
\end{figure}

\subsection{Quickest abnormal event detection with zero-bias DNNs}
Our model can detect abnormalities (the appearance of an unknown aircraft's signal) with almost neglectable latency (less than ten samples on average using GLR chart) as a result of both high true-positive and true-negative rates. To further evaluate our proposed method, we can use numerical simulation results to demonstrate the performance of zero-bias DNN, since its response characteristics prior and after abnormal events are modeled as two Bernoulli distributions in Section~\ref{sectBernoulliFormalization}. We experiment with a collection of possible values of $h_{GLR}$, $FPR$, and $TPR$ that a zero-bias abnormality detector can encounter. In which $TPR \in [0.6,0.99]$, $FPR \in[0,0.4]$ and $h_{GLR} \in [2,20]$

\begin{figure}[h]
\centering
\includegraphics[width=0.85\linewidth]{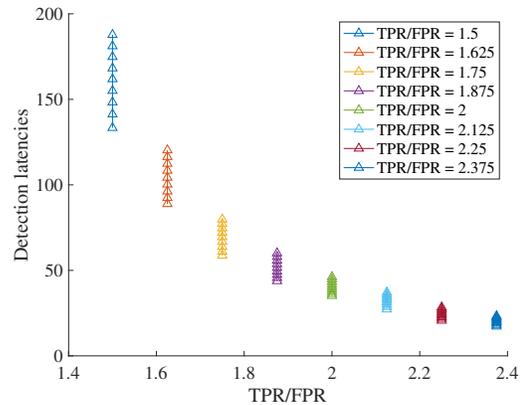}
\caption{Distribution of abnormal event detection latencies.}
\label{figLatencyVsTPRRatio}
\end{figure}

The relationship between abnormal event detection latency and false alarm rate is depicted in Figure~\ref{figLatencyVsARL}. We discover a nice cut-off property, in which the average false alarm rate becomes zero as we select a proper detection threshold $h_{GLR}$ or $h_{CUSUM}$. After the threshold is properly set. Once the detection threshold is greater than a certain value, 12 in our experiment, the detection delay grows linearly as depicted in Figure~\ref{figLatencyVsThresholds}. A further analysis of the distribution of detection delay with threshold $h_{GLR} \in [12, 20]$ is presented in Figure~\ref{figLatencyVsTPRRatio}, in general, as $TPR/FPR$ gets larger, average detection latency decreases with less sensitive to $h_{GLR}$.

\section{Conclusion}
\label{sectCC}
In this paper, we significantly extend the analysis of our previously proposed zero-bias DNN and combine it with the Quickest Event Detection algorithms to detect abnormalities and time-dependent abnormal events in IoT with the lowest assured latency. We first analyze the zero-bias dense layer and provide to use Voronoi diagram to increase the explainability of DNN models. We then provide a solution to convert zero-bias DNN classifiers, which are easier to obtain, into performance assured binary abnormality detectors with assured performance boundaries. Using the converted abnormality detectors, we model their behaviors using Bernoulli distribution, which perfectly adapts to the Generalized Likelihood Ratio based Quickest Detection scheme. In this Quickest Detection scheme, the theoretically assured lowest abnormal event detection delay is provided with predictable false alarms. Finally, we demonstrate the framework's effectiveness using both massive signal records from real-world aviation communication systems and simulated data.

In the future, we will investigate the possibility of using similar methods to visualize conventional deep neural networks' decision boundaries. We only use the Voronoi diagram in the 2D plane in this research, and better data dimension reduction methods and visualization methods specifically focus on the hyperspherical plane would be proposed.

\section*{Acknowledgment}

This research was partially supported by the National Science Foundation under Grant No. 1956193.




\bibliographystyle{IEEEtran}

\bibliography{ReviewRef.bib}

\begin{thebibliography}{10}
\providecommand{\url}[1]{#1}
\csname url@samestyle\endcsname
\providecommand{\newblock}{\relax}
\providecommand{\bibinfo}[2]{#2}
\providecommand{\BIBentrySTDinterwordspacing}{\spaceskip=0pt\relax}
\providecommand{\BIBentryALTinterwordstretchfactor}{4}
\providecommand{\BIBentryALTinterwordspacing}{\spaceskip=\fontdimen2\font plus
\BIBentryALTinterwordstretchfactor\fontdimen3\font minus
  \fontdimen4\font\relax}
\providecommand{\BIBforeignlanguage}[2]{{%
\expandafter\ifx\csname l@#1\endcsname\relax
\typeout{** WARNING: IEEEtran.bst: No hyphenation pattern has been}%
\typeout{** loaded for the language `#1'. Using the pattern for}%
\typeout{** the default language instead.}%
\else
\language=\csname l@#1\endcsname
\fi
#2}}
\providecommand{\BIBdecl}{\relax}
\BIBdecl

\bibitem{xu2020rf}
C.~Xu, B.~Chen, Y.~Liu, F.~He, and H.~Song, ``Rf fingerprint measurement for
  detecting multiple amateur drones based on stft and feature reduction,'' in
  \emph{2020 Integrated Communications Navigation and Surveillance Conference
  (ICNS)}.\hskip 1em plus 0.5em minus 0.4em\relax IEEE, 2020, pp. 4G1--1.

\bibitem{jiang2020applying}
Y.~Jiang, Y.~Liu, D.~Liu, and H.~Song, ``Applying machine learning to aviation
  big data for flight delay prediction,'' in \emph{2020 IEEE
  DASC/PiCom/CBDCom/CyberSciTech}.\hskip 1em plus 0.5em minus 0.4em\relax IEEE,
  2020, pp. 665--672.

\bibitem{wang2020dynamic}
L.~Wang, X.~Yue, H.~Wang, K.~Ling, Y.~Liu, J.~Wang, J.~Hong, W.~Pen, and
  H.~Song, ``Dynamic inversion of inland aquaculture water quality based on
  uavs-wsn spectral analysis,'' \emph{Remote Sensing}, vol.~12, no.~3, p. 402,
  2020.

\bibitem{peng2018modulation}
S.~Peng, H.~Jiang, H.~Wang, H.~Alwageed, Y.~Zhou, M.~M. Sebdani, and Y.-D. Yao,
  ``Modulation classification based on signal constellation diagrams and deep
  learning,'' \emph{IEEE transactions on neural networks and learning systems},
  vol.~30, no.~3, pp. 718--727, 2018.

\bibitem{gao2019eeg}
Z.~Gao, X.~Wang, Y.~Yang, C.~Mu, Q.~Cai, W.~Dang, and S.~Zuo, ``Eeg-based
  spatio--temporal convolutional neural network for driver fatigue
  evaluation,'' \emph{IEEE transactions on neural networks and learning
  systems}, vol.~30, no.~9, pp. 2755--2763, 2019.

\bibitem{jiang2019uncertainty}
Y.~Jiang, M.~Wang, X.~Jiao, H.~Song, H.~Kong, R.~Wang, Y.~Liu, J.~Wang, and
  J.~Sun, ``Uncertainty theory based reliability-centric cyber-physical system
  design,'' in \emph{2019 International Conference on Internet of Things
  (iThings) and IEEE GreenCom/CPSCom/SmartData}.\hskip 1em plus 0.5em minus
  0.4em\relax IEEE, 2019, pp. 208--215.

\bibitem{perera2017efficient}
P.~Perera and V.~M. Patel, ``Efficient and low latency detection of intruders
  in mobile active authentication,'' \emph{IEEE Transactions on Information
  Forensics and Security}, vol.~13, no.~6, pp. 1392--1405, 2017.

\bibitem{das2020opportunities}
A.~Das and P.~Rad, ``Opportunities and challenges in explainable artificial
  intelligence (xai): A survey,'' \emph{arXiv preprint arXiv:2006.11371}, 2020.

\bibitem{kirkpatrick2017overcoming}
J.~Kirkpatrick, R.~Pascanu, N.~Rabinowitz, J.~Veness, G.~Desjardins, A.~A.
  Rusu, K.~Milan, J.~Quan, T.~Ramalho, A.~Grabska-Barwinska \emph{et~al.},
  ``Overcoming catastrophic forgetting in neural networks,'' \emph{Proceedings
  of the national academy of sciences}, vol. 114, no.~13, pp. 3521--3526, 2017.

\bibitem{rolnick2019experience}
D.~Rolnick, A.~Ahuja, J.~Schwarz, T.~Lillicrap, and G.~Wayne, ``Experience
  replay for continual learning,'' in \emph{Advances in Neural Information
  Processing Systems}, 2019, pp. 350--360.

\bibitem{liu2020zero}
Y.~Liu, J.~Wang, J.~Li, H.~Song, T.~Yang, S.~Niu, and Z.~Ming, ``Zero-bias deep
  learning for accurate identification of internet of things (iot) devices,''
  \emph{IEEE Internet of Things Journal}, 2020.

\bibitem{wong2018clustering}
L.~J. Wong, W.~C. Headley, S.~Andrews, R.~M. Gerdes, and A.~J. Michaels,
  ``Clustering learned cnn features from raw i/q data for emitter
  identification,'' in \emph{MILCOM 2018-2018 IEEE Military Communications
  Conference (MILCOM)}.\hskip 1em plus 0.5em minus 0.4em\relax IEEE, 2018, pp.
  26--33.

\bibitem{scheirer2012toward}
W.~J. Scheirer, A.~de~Rezende~Rocha, A.~Sapkota, and T.~E. Boult, ``Toward open
  set recognition,'' \emph{IEEE transactions on pattern analysis and machine
  intelligence}, vol.~35, no.~7, pp. 1757--1772, 2012.

\bibitem{bendale2016towards}
A.~Bendale and T.~E. Boult, ``Towards open set deep networks,'' in
  \emph{Proceedings of the IEEE conference on computer vision and pattern
  recognition}, 2016, pp. 1563--1572.

\bibitem{roy2019rfal}
D.~Roy, T.~Mukherjee, M.~Chatterjee, E.~Blasch, and E.~Pasiliao, ``Rfal:
  Adversarial learning for rf transmitter identification and classification,''
  \emph{IEEE Transactions on Cognitive Communications and Networking}, 2019.

\bibitem{shi2019deep}
Y.~Shi, K.~Davaslioglu, Y.~E. Sagduyu, W.~C. Headley, M.~Fowler, and G.~Green,
  ``Deep learning for rf signal classification in unknown and dynamic spectrum
  environments,'' in \emph{2019 IEEE International Symposium on Dynamic
  Spectrum Access Networks (DySPAN)}.\hskip 1em plus 0.5em minus 0.4em\relax
  IEEE, 2019, pp. 1--10.

\bibitem{lai2008quickest}
L.~Lai, Y.~Fan, and H.~V. Poor, ``Quickest detection in cognitive radio: A
  sequential change detection framework,'' in \emph{IEEE GLOBECOM 2008-2008
  IEEE Global Telecommunications Conference}.\hskip 1em plus 0.5em minus
  0.4em\relax IEEE, 2008, pp. 1--5.

\bibitem{poor2008quickest}
H.~V. Poor and O.~Hadjiliadis, \emph{Quickest detection}.\hskip 1em plus 0.5em
  minus 0.4em\relax Cambridge University Press, 2008.

\bibitem{johnson2017detecting}
P.~Johnson, J.~Moriarty, and G.~Peskir, ``Detecting changes in real-time data:
  a user’s guide to optimal detection,'' \emph{Philosophical Transactions of
  the Royal Society A: Mathematical, Physical and Engineering Sciences}, vol.
  375, no. 2100, p. 20160298, 2017.

\bibitem{basseville1993detection}
M.~Basseville, I.~V. Nikiforov \emph{et~al.}, \emph{Detection of abrupt
  changes: theory and application}.\hskip 1em plus 0.5em minus 0.4em\relax
  prentice Hall Englewood Cliffs, 1993, vol. 104.

\bibitem{liu2020deep}
Y.~Liu, J.~Wang, S.~Niu, and H.~Song, ``Deep learning enabled reliable identity
  verification and spoofing detection,'' in \emph{International Conference on
  Wireless Algorithms, Systems, and Applications}.\hskip 1em plus 0.5em minus
  0.4em\relax Springer, 2020, pp. 333--345.

\bibitem{erwig2000graph}
M.~Erwig, ``The graph voronoi diagram with applications,'' \emph{Networks: An
  International Journal}, vol.~36, no.~3, pp. 156--163, 2000.

\bibitem{matlabMinst}
MathWorks, ``Create simple deep learning network for classification,''
  \url{https://www.mathworks.com/help/deeplearning/ug/create-simple-deep-learning-network-for-classification.html},
  May 2018.

\bibitem{maaten2008visualizing}
L.~v.~d. Maaten and G.~Hinton, ``Visualizing data using t-sne,'' \emph{Journal
  of machine learning research}, vol.~9, no. Nov, pp. 2579--2605, 2008.

\bibitem{choi2009least}
Y.-S. Choi, ``Least squares one-class support vector machine,'' \emph{Pattern
  Recognition Letters}, vol.~30, no.~13, pp. 1236--1240, 2009.

\bibitem{de2000mahalanobis}
R.~De~Maesschalck, D.~Jouan-Rimbaud, and D.~L. Massart, ``The mahalanobis
  distance,'' \emph{Chemometrics and intelligent laboratory systems}, vol.~50,
  no.~1, pp. 1--18, 2000.

\bibitem{breunig2000lof}
M.~M. Breunig, H.-P. Kriegel, R.~T. Ng, and J.~Sander, ``Lof: identifying
  density-based local outliers,'' in \emph{Proceedings of the 2000 ACM SIGMOD
  international conference on Management of data}, 2000, pp. 93--104.

\bibitem{hypersphere}
G.~Dorini, ``n-dimensional hypersphere in parametric coordinates. \text{MATLAB
  Central File Exchange. Retrieved November 25, 2020.}''
  \url{https://www.mathworks.com/matlabcentral/fileexchange/5397-hypersphere}.

\bibitem{weisstein2002bernoulli}
E.~W. Weisstein, ``Bernoulli distribution,'' \emph{https://mathworld. wolfram.
  com/}, 2002.

\bibitem{huang2013generalized}
W.~Huang, S.~Wang, and M.~R. Reynolds~Jr, ``A generalized likelihood ratio
  chart for monitoring bernoulli processes,'' \emph{Quality and Reliability
  Engineering International}, vol.~29, no.~5, pp. 665--679, 2013.

\bibitem{lai1998information}
T.~L. Lai, ``Information bounds and quick detection of parameter changes in
  stochastic systems,'' \emph{IEEE Transactions on Information Theory},
  vol.~44, no.~7, pp. 2917--2929, 1998.

\bibitem{huang2012binomial}
W.~Huang, M.~R. Reynolds~Jr, and S.~Wang, ``A binomial glr control chart for
  monitoring a proportion,'' \emph{Journal of Quality Technology}, vol.~44,
  no.~3, pp. 192--208, 2012.

\bibitem{yongxin_liu_2020_4248678}
\BIBentryALTinterwordspacing
L.~Yongxin, ``Adabelief-matlab (github),'' Nov. 2020. [Online]. Available:
  \url{https://doi.org/10.5281/zenodo.4248678}
\BIBentrySTDinterwordspacing

\bibitem{gt9v-kz32-20}
\BIBentryALTinterwordspacing
Y.~Liu, J.~Wang, H.~Song, S.~Niu, and Y.~Thomas, ``A 24-hour signal recording
  dataset with labels for cybersecurity and \text{IoT},'' 2020. [Online].
  Available: \url{http://dx.doi.org/10.21227/gt9v-kz32}
\BIBentrySTDinterwordspacing

\bibitem{riddle1090}
J.~Sun, ``An open-access book about decoding mode-s and ads-b data,''
  \url{https://mode-s.org/}, May 2017.

\end{thebibliography}
%

%
\begin{IEEEbiography}[{\includegraphics[width=1in,height=1.25in,clip,keepaspectratio]{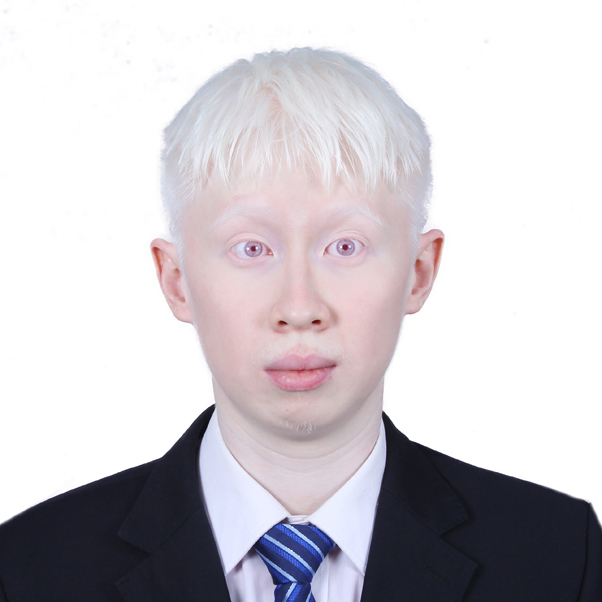}}]{YongXin Liu}
(LIU11@my.erau.edu) received his first Ph.D. from South China University of Technology (SCUT) and currently working towards his second Ph.D. in the Department of Electrical Engineering and Computer Science, Embry-Riddle Aeronautical University, Daytona Beach, FL. His major research interests include data mining, wireless networks, the Internet of Things, and unmanned aerial vehicles. 
\end{IEEEbiography}

\begin{IEEEbiography}[{\includegraphics[width=1in,height=1.25in,clip,keepaspectratio]{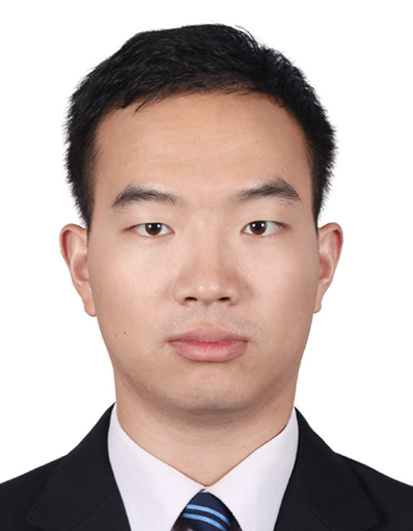}}]{Jian Wang}
(wangj14@my.erau.edu) is a Ph.D. student in the Department of Electrical Engineering and Computer Science, Embry-Riddle Aeronautical University (ERAU), Daytona Beach, Florida, and a graduate research assistant in the Security and Optimization for Networked Globe Laboratory (SONG Lab, www.SONGLab.us). He received his M.S. from South China Agricultural University (SCAU) in 2017. His research interests include wireless networks, unmanned aerial systems, and machine learning.
\end{IEEEbiography}
\begin{IEEEbiography}[{\includegraphics[width=1in,height=1.25in,clip,keepaspectratio]{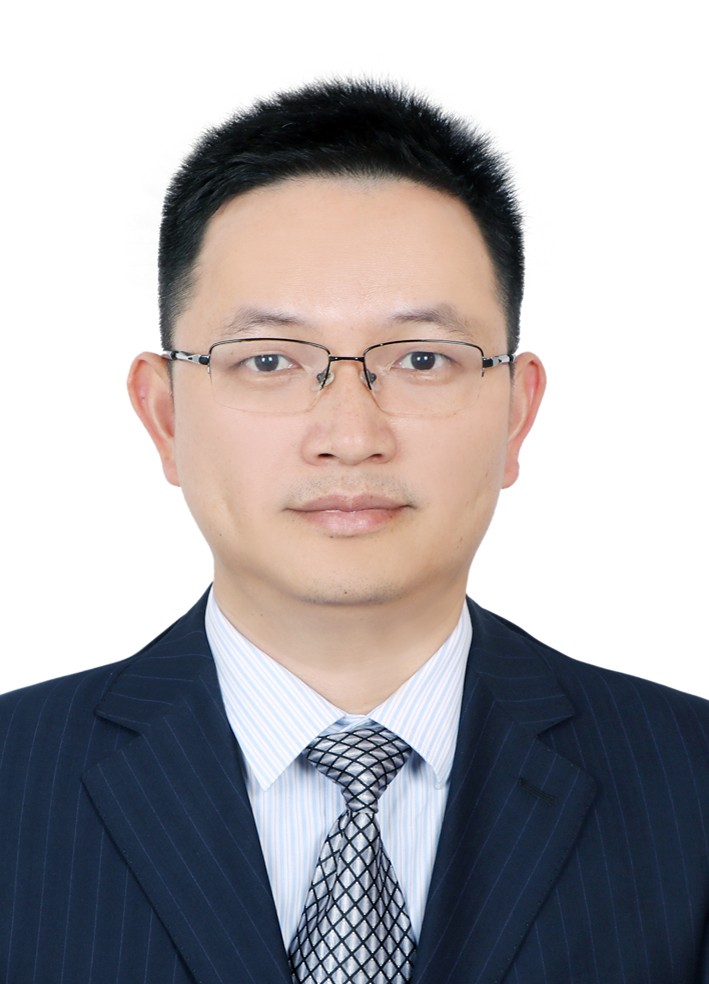}}]{Jianqiang Li}
(lijq@szu.edu.cn) received his B.S. and Ph.D.
degrees from the South China University of
Technology in 2003 and 2008, respectively. He is a Professor with the College of Computer
and Software Engineering, Shenzhen University,
Shenzhen, China. His major
research interests include Internet of Things, robotic,
hybrid systems, and embedded systems.
\end{IEEEbiography}

\begin{IEEEbiography}[{\includegraphics[width=1in,height=1.1in,clip,keepaspectratio]{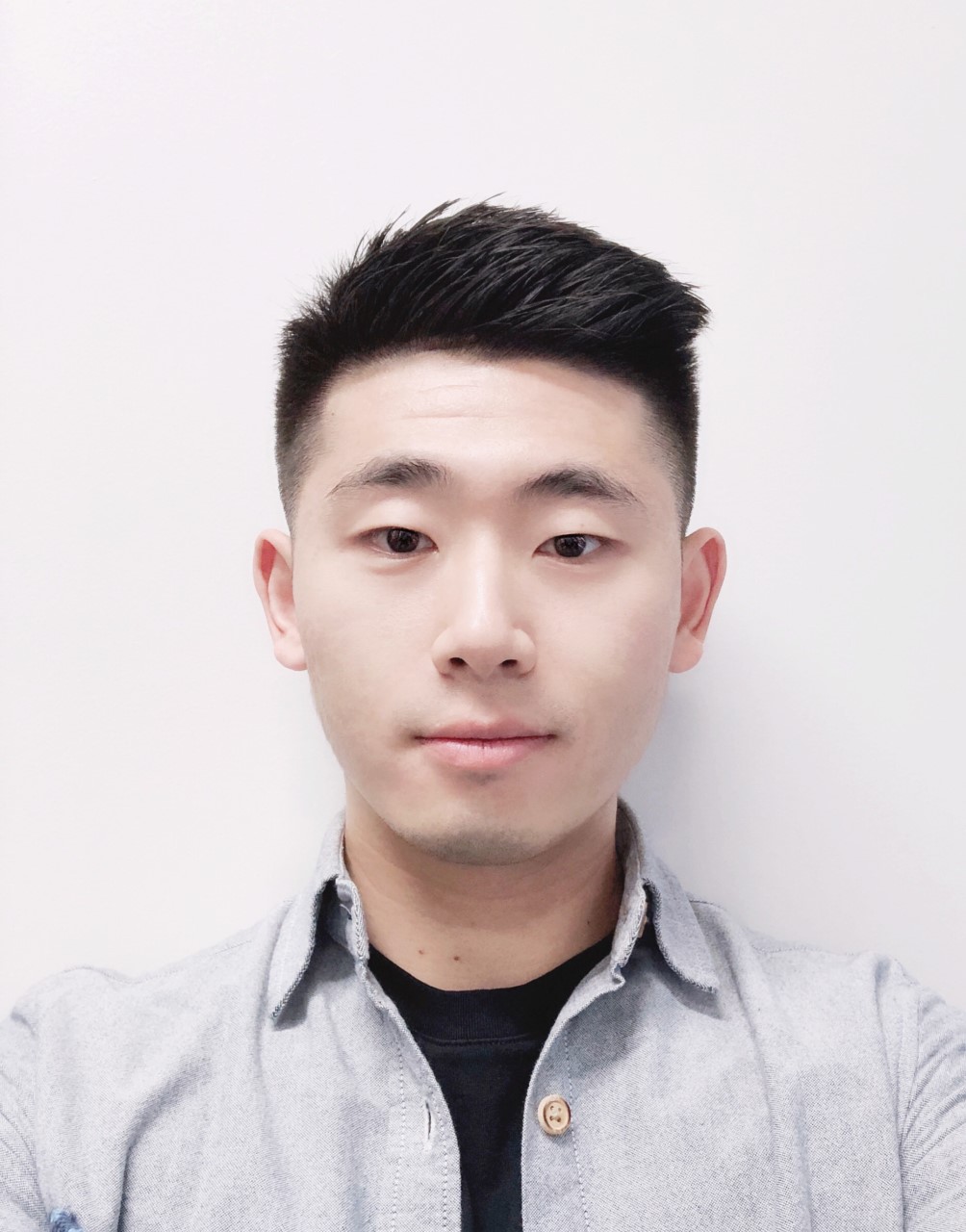}}]{Shuteng Niu}
(shutengn@my.erau.edu) is a Ph.D. student in the Department of Electrical Engineering and Computer Science, Embry-Riddle Aeronautical University (ERAU), Daytona Beach, Florida, and a graduate research assistant in the Security and Optimization for Networked Globe Laboratory (SONG Lab, www.SONGLab.us). He received his M.S. from ERAU in 2018. His research interests include machine learning, data mining, and signal processing.
\end{IEEEbiography}
\vspace{-6em}
\begin{IEEEbiography}[{\includegraphics[width=1in,height=1.25in,clip,keepaspectratio]{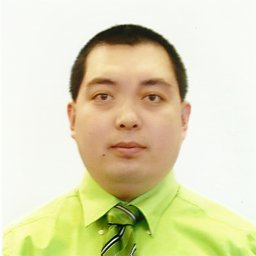}}]{Houbing Song} (M'12–SM'14) received the Ph.D. degree in electrical engineering from the University of Virginia, Charlottesville, VA, in August 2012, and the M.S. degree in civil engineering from the University of Texas, El Paso, TX, in December 2006.

In August 2017, he joined the Department of Electrical Engineering and Computer Science, Embry-Riddle Aeronautical University, Daytona Beach, FL, where he is currently an Assistant Professor and the Director of the Security and Optimization for Networked Globe Laboratory (SONG Lab, www.SONGLab.us). He served on the faculty of West Virginia University from August 2012 to August 2017. In 2007 he was an Engineering Research Associate with the Texas A\&M Transportation Institute. He has served as an Associate Technical Editor for IEEE Communications Magazine (2017-present), an Associate Editor for IEEE Internet of Things Journal (2020-present) and IEEE Journal on Miniaturization for Air and Space Systems (J-MASS) (2020-present), and a Guest Editor for IEEE Journal on Selected Areas in Communications (J-SAC), IEEE Internet of Things Journal, IEEE Transactions on Industrial Informatics, IEEE Sensors Journal, IEEE Transactions on Intelligent Transportation Systems, and IEEE Network. He is the editor of six books, including Big Data Analytics for Cyber-Physical Systems: Machine Learning for the Internet of Things, Elsevier, 2019,  Smart Cities: Foundations, Principles and Applications, Hoboken, NJ: Wiley, 2017, Security and Privacy in Cyber-Physical Systems: Foundations, Principles and Applications, Chichester, UK: Wiley-IEEE Press, 2017, Cyber-Physical Systems: Foundations, Principles and Applications, Boston, MA: Academic Press, 2016, and Industrial Internet of Things: Cybermanufacturing Systems, Cham, Switzerland: Springer, 2016.  He is the author of more than 100 articles. His research interests include cyber-physical systems, cybersecurity and privacy, internet of things, edge computing, AI/machine learning, big data analytics, unmanned aircraft systems, connected vehicle, smart and connected health, and wireless communications and networking. His research has been featured by popular news media outlets, including IEEE GlobalSpec's Engineering360, USA Today, U.S. News \& World Report, Fox News, Association for Unmanned Vehicle Systems International (AUVSI), Forbes, WFTV, and New Atlas.

Dr. Song is a senior member of ACM and an ACM Distinguished Speaker. Dr. Song was a recipient of the Best Paper Award from the 12th IEEE International Conference on Cyber, Physical and Social Computing (CPSCom-2019), the Best Paper Award from the 2nd IEEE International Conference on Industrial Internet (ICII 2019), the Best Paper Award from the 19th Integrated Communication, Navigation and Surveillance technologies (ICNS 2019) Conference, the Best Paper Award from the 6th IEEE International Conference on Cloud and Big Data Computing (CBDCom 2020), and the Best Paper Award from the 15th International Conference on Wireless Algorithms, Systems, and Applications (WASA 2020).

\end{IEEEbiography}




\end{document}